\documentclass{PoS}
\usepackage{graphicx}

\title{Fits of $SU(3)$ $N_f=8$ data to dilaton-pion effective field theory}

\ShortTitle{Fits to dilaton-pion effective field theory}

\author{\speaker{Maarten Golterman}\\
        Department of Physics and Astronomy, San Francisco State University\\ San Francisco, CA 94132, USA\\
        E-mail: \email{maarten@sfsu.edu}}

\author{Yigal Shamir\\
        Raymond and Beverly Sackler School of Physics and Astronomy, Tel Aviv University\\ 69978 Tel Aviv, Israel\\
        E-mail: \email{shamir@post.tau.ac.il}}

\abstract{We report on fits of the $SU(3)$ $N_f=8$ LSD spectral data to chiral perturbation theory with a dilatonic meson. These fits confirm that current simulations are in the ``large-mass'' regime, with approximate hyperscaling as the leading mass dependence. We find that the leading-order effective field theory describes the data well. In particular, the effective field theory allows us to understand the staggered taste splitting, explaining the pattern observed in the LSD data, which looks different from QCD.}

\FullConference{37th International Symposium on Lattice Field Theory - Lattice2019\\
		16-22 June 2019\\
		Wuhan, China}

\begin{document}

\section{Introduction}
Numerical simulations of the $SU(3)$ gauge theory with $N_f=8$ flavors of fundamental
fermions show that the spectrum of the theory behaves quite differently from the spectrum
of a similar theory with far fewer (light) flavors, such as QCD \cite{LSD1,LSD2}
(see also Ref.~\cite{LatKMI}).   The salient differences are
three-fold:  (1)  the 8-flavor theory contains a stable flavor-singlet $0^{++}$ state which, at the fermion
masses explored in the simulations, is light and approximately degenerate with the 
pseudo-scalar Nambu--Goldstone (NG) bosons associated with chiral symmetry breaking
(``pions''),
(2)  dimensionless ratios of hadronic quantities are nearly independent of the fermion mass (over a
range of fermion masses differing by factors up to 7), and (3) a taste-breaking pattern 
that looks very different from that in QCD (staggered fermions were used for the 
simulations of Refs.~\cite{LSD1,LSD2}).

While the simulations indicate that the 8-flavor theory breaks chiral symmetry also in the
chiral limit, with the light $0^{++}$ state in the spectrum standard chiral 
perturbation theory (ChPT) clearly cannot be used as a low-energy effective theory.   Instead, it 
should be expanded to include the physics of the light scalar state, and, if this is to be
done systematically, a guiding principle that leads to a power counting scheme is 
required.   Such an effective theory was proposed based on the assumption that 
the light scalar is an approximate NG boson for the breaking of scale invariance,
which is assumed to be small because of the proximity of the 8-flavor theory to the
conformal window.   In Refs.~\cite{GS1,GS2,GS3,GS4}, it is assumed that the difference
of the number of flavors, $N_f$, with the critical value $N_f^*(N_c)$ at which theory enters
the conformal phase in which the theory develops an infra-red fixed point (IRFP)
can be used as an expansion parameter.   More precisely, the small parameter
is $n_f-n_f^*$, with $n_f=\lim_{N_c\to\infty} N_f/N_c$
and $n_f^*=\lim_{N_c\to\infty} N_f^*(N_c)/N_c$, where the Veneziano limit
$N_c,\ N_f\to\infty$ with fixed ratio is taken.   As explained in detail in Refs.~\cite{GS1,GS2},
this assumption allows us to augment standard ChPT with an effective field describing the
light scalar, which we will refer to as the dilatonic meson, or dilaton.\footnote{For more
on the assumptions underlying our framework, see Ref.~\cite{GS4}.}   
We will refer to this extension of ChPT as dChPT.

Here, we test tree-level dChPT on the published data of Ref.~\cite{LSD2}, as a natural
first step.  (With the currently attained numerical precision, NLO effects are 
unlikely to be quantitatively accessible.)   Some tree-level tests have been carried
out in Ref.~\cite{LatHC}; the results we report below are in agreement with those
reported in Ref.~\cite{LatHC}.   A new result is our exploration of taste-breaking
effects, for which we refer to Sec.~3 below.   We emphasize that all results reported
here are based on fits of the data as published in Ref.~\cite{LSD2}.   No correlations
have been taken into account, and all results should be considered preliminary.
Work on a more complete analysis of the numerical data is in progress \cite{GNS}.

\section{dChPT at tree level}
The leading-order dChPT lagrangian is \cite{GS1}
\begin{equation}
\label{lag}
{\cal L}=\frac{1}{4}{\hat f}_\pi^2 e^{2\tau}{\rm tr}(\partial_\mu\Sigma^\dagger\partial_\mu\Sigma)
+\frac{1}{2}{\hat f}_\tau^2 e^{2\tau}\partial_\mu\tau\partial_\mu\tau-\frac{1}{2}{\hat f}_\pi^2{\hat B}_\pi e^{(3-\gamma_*)\tau}m\,{\rm tr}(\Sigma+\Sigma^\dagger)
+{\hat f}_\tau^2{\hat B}_\tau e^{4\tau}c_1\left(\tau-\frac{1}{4}\right)\ .
\end{equation}
Here $\tau$ is the field describing the dilatonic meson, and $\Sigma={\rm exp}(2i\pi/{\hat f}_\pi)$ 
is the usual field describing the pions; ${\hat f}_\pi$, ${\hat f}_\tau$, ${\hat B}_\pi$, and ${\hat B}_\tau$
are low-energy constants, and $\gamma_*$ is the value of the 
mass-anomalous dimension at the nearby IRFP \cite{GS1,GS3}.   The small parameters are
the fermion mass $m>0$, and $c_1\propto n_f-n_f^*$.  At fixed $n_f$, the dilatonic meson $\tau$ decouples in the $m\to 0$ limit, in which the
pions are described by ordinary ChPT (for 8 light flavors).
For larger values of $m$, there exists a ``large-mass'' regime 
which exhibits approximate hyper-scaling; dChPT is applicable as long as  $c_1\log{m}\ll 1$ \cite{GS4}.
   The $\tau$ field has been 
shifted such that $v\equiv\langle\tau\rangle=0$ for $m=0$.

First, the classical potential is minimized by setting $\Sigma=1$, and solving 
\begin{equation}
\label{minpot}
\frac{m}{c_1{\hat M}}=v(m)e^{(1+\gamma_*)v(m)}\ ,\qquad{\hat M}=\frac{4{\hat f}_\tau^2{\hat B}_\tau}{{\hat f}_\pi^2{\hat B}_\pi N_f(3-\gamma_*)}\ ,
\end{equation}
for $v(m)$.   Note that this is an $O(1)$ relation, because both $m$ and $c_1$ are small, and
assumed to be of the same order.   Some of the tree-level predictions following from Eqs.~(\ref{lag})
and (\ref{minpot}) are
\begin{eqnarray}
\label{treelevel}
F_{\pi,\tau}&=&{\hat f}_{\pi,\tau}e^{v(m)}\ ,\\
M_\pi^2&=&2{\hat B}_\pi m e^{(1-\gamma_*)v(m)}\ ,\nonumber\\
M_\tau^2&=&4c_1{\hat B}_\tau e^{2v(m)}\left(1+(1+\gamma_*)v(m)\right)\ .\nonumber
\end{eqnarray}
These results can be combined into the relations
\begin{eqnarray}
\label{treerel}
M_\pi^2 F_\pi^{\gamma_*-1}&=&Cm\ ,\\
\frac{m}{F_\pi}&=&D_2\,\frac{M_\pi^2}{F_\pi^2}\,{\rm exp}\left(D_1\frac{M_\pi^2}{F_\pi^2}\right)\ ,\nonumber\\
{\hat f}_\pi&=&(CD_2)^{1/\gamma_*}\ ,\nonumber\\
v(m)&=&\frac{D_1}{\gamma_*}\,\frac{M_\pi^2}{F_\pi^2}\ ,\nonumber
\end{eqnarray}
where $C$ and $D_{1,2}$ are combinations of low-energy constants, including $\gamma_*$.

Assuming that the lattice spacing $a$ does not depend on $m$, we can fit the first two
relations (the second relation is in terms of dimensionless ratios, so the assumption is not
needed in that case).    Preliminary results are, in units of the lattice spacing\footnote{Early results indicate that proper
correlated fits lead to results consistent with Eq.~(\ref{fitresults}) with a good fit
quality \cite{GNS}.}
\begin{eqnarray}
\label{fitresults}
\gamma_*&=&0.936(19)\ ,\qquad C=1.93(6)\ ,\\
D_1&=&0.22(3)\ ,\qquad \log{D_2}=-8.8(5)\ .\nonumber
\end{eqnarray}
From the last line of Eq.~(\ref{treerel}) we then find that $a{\hat f}_\pi\sim 0.0006$.    This is 
much smaller than the computed values of $aF_\pi$, which range from $0.021$ to $0.053$
for fermion masses between $am=0.00125$ and $am=0.00889$.   This large difference is
explained by the factor $e^{v(m)}$ in the first line of Eq.~(\ref{treelevel}).   These factors
thus range between about $35$ and $90$, indicating that the numerical simulations of
Refs.~\cite{LSD1,LSD2} are in the ``large-mass'' regime \cite{GS4}, where the factor
$e^{(1+\gamma_*)v(m)}$ dominates over the factor $v(m)$ in Eq.~(\ref{minpot}).   In this
regime, the theory exhibits approximate hyperscaling.    Figure~1 shows that results for 
the ratios of masses and $F_\pi$ obtained by Ref.~\cite{LSD2} are consistent
with approximate hyperscaling as predicted by dChPT.  Needless
to say, this behavior is quite different from that in QCD.
\begin{figure}[t]
\center{\includegraphics[width=12cm] {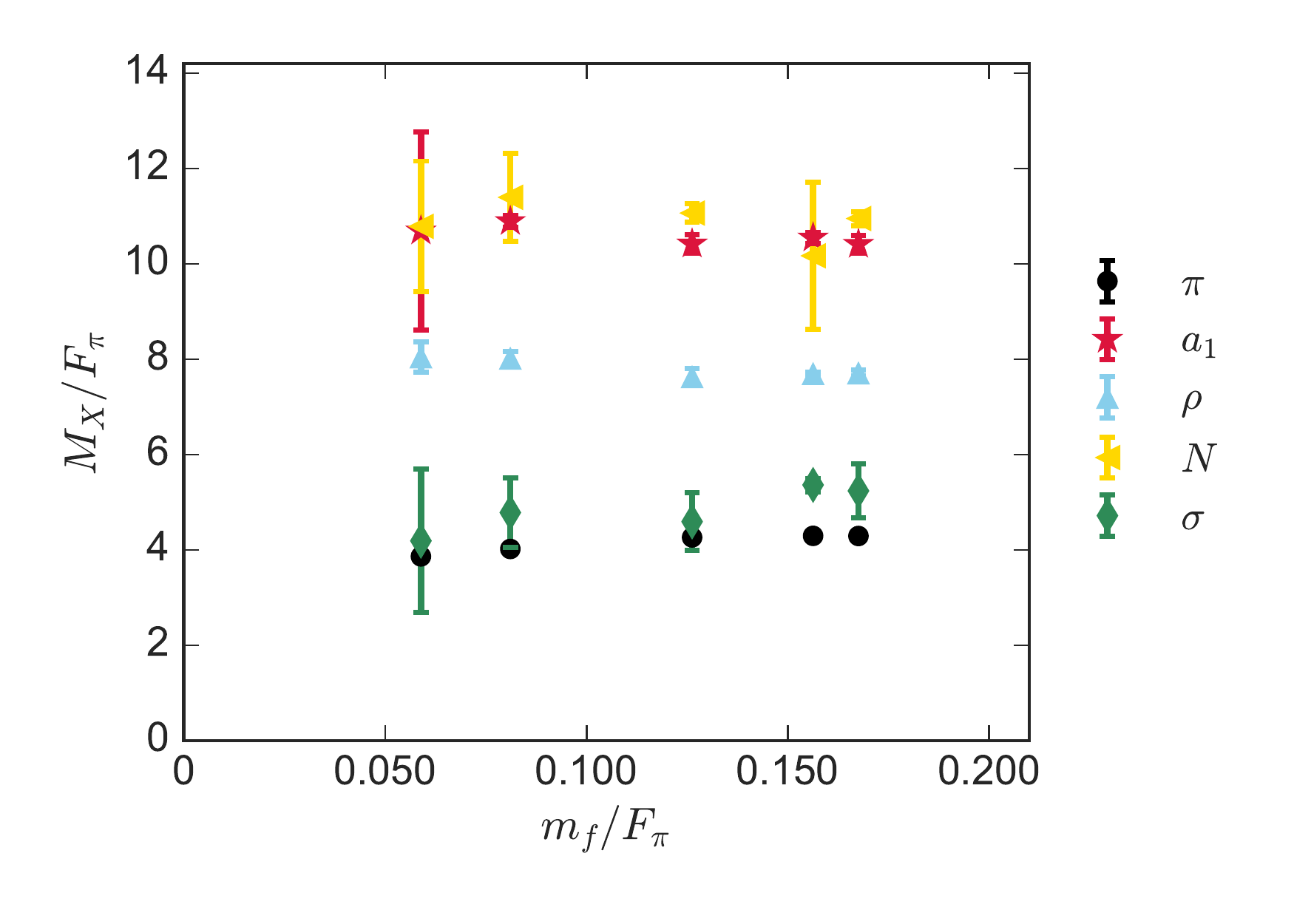}}
\caption{Ratios of hadron masses over $F_\pi$, as a function of the fermion mass
in units of $F_\pi$, from Ref.~\cite{LSD2}.  (Figure courtesy of E.~Neil.)}
\end{figure}
Since ${\hat B}_\pi$ is by construction independent of $m$, we can test the 
independence of the lattice spacing by
computing values for the low-energy constant $a{\hat B}_\pi$ in lattice
units.  From the fourth line of Eq.~(\ref{treerel}) one obtains $v(m)$, and then
from the second line of Eq.~(\ref{treelevel}) one computes $a{\hat B}_\pi$.  
Doing this, we find that this quantity is constant to within 3\% over the range of fermion
masses considered in the numerical simulations, which is well within the errors on the
computed values at each $am$.   This validates our assumption
that the lattice spacing does not depend on $m$.

\section{Taste breaking}
The simulations of Refs.~\cite{LSD1,LSD2} were performed with staggered fermions,
and Ref.~\cite{LSD2} reported, in addition to the ``exact'' NG pion mass $M_\pi$ also
the values for the ``taste'' pions $M_{i5}$ and $M_{ij}$.  Mass-squared differences are shown in Fig.~2. 
Both taste splittings show a strong dependence on the fermion mass at fixed lattice
spacing.   This is very much unlike QCD, where, if one plots the differences between
the squares of the masses of the different tastes, one would find virtually no dependence
on the fermion mass, {\it i.e.}, horizontal lines (see, for example, Fig.~3 in Ref.~\cite{MGLH}).

An important question is whether the dChPT framework can explain this salient
difference, which has also been observed in the $SU(3)$ theory with two sextet fermions
\cite{LatHC2}.
In staggered QCD, taste splittings
can be understood in terms of staggered ChPT \cite{LS,AB} (for reviews,
see Refs.~\cite{MGLH,MILC}).
\begin{figure}[t]
\center{\includegraphics[width=12cm] {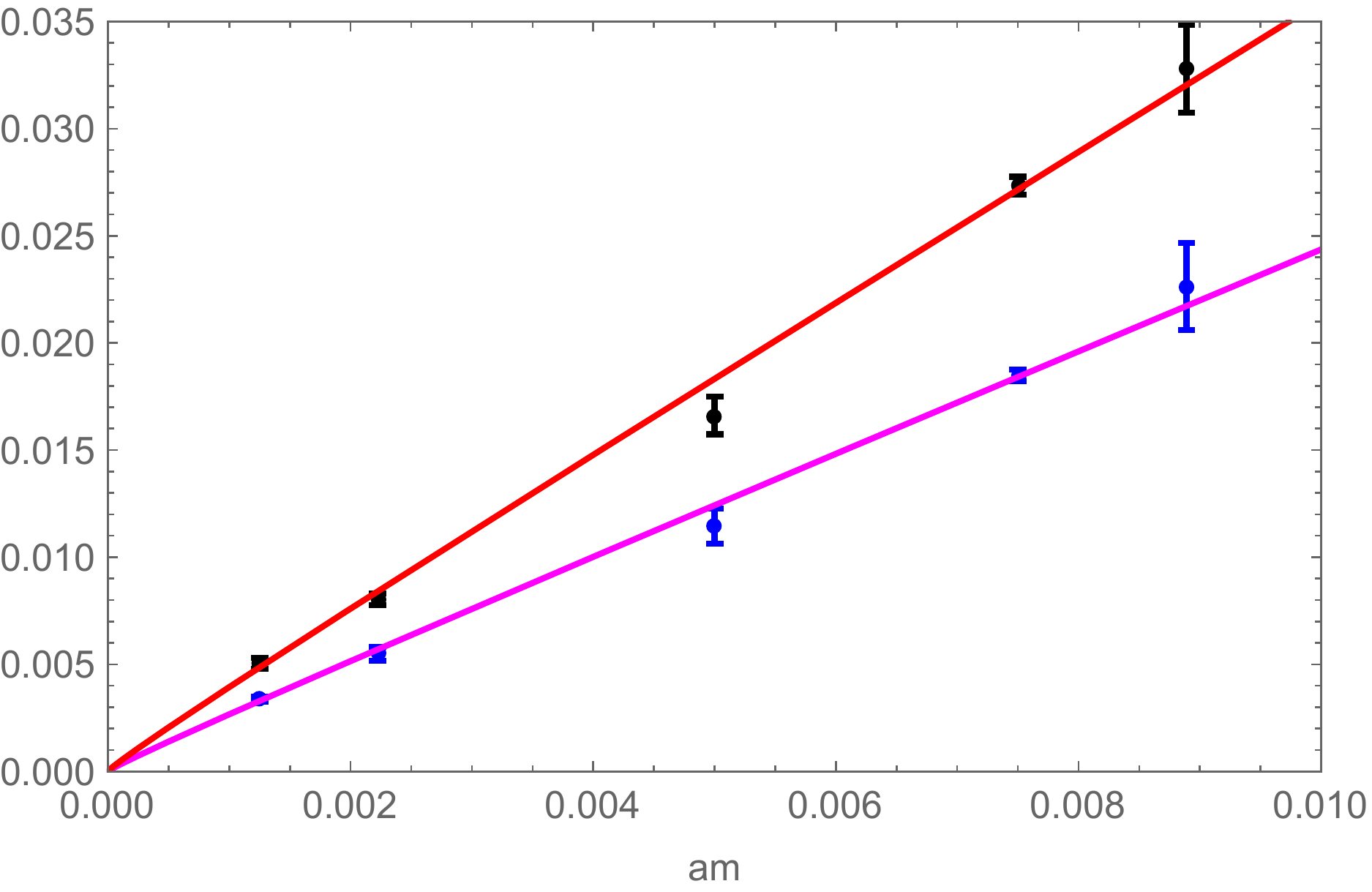}}
\caption{Taste splittings $(aM_{ij})^2-(aM_\pi)^2$ (black points; red curve), and
$(aM_{i5})^2-(aM_\pi)^2$ (blue points; magenta curve), as a function of $am$.
Data from Ref.~\cite{LSD2}.}
\end{figure}
In the Symanzik effective action, the leading-order taste-breaking effects are
encapsulated by four-fermion operators of the form \cite{LS}
\begin{equation}
\label{SET}
a^2(\overline\psi\Gamma\psi)(\overline\psi\Gamma\psi)\ ,
\end{equation}
where $\Gamma$ is a gamma-matrix acting on the taste index of $\psi$.
The operator $(\overline\psi\Gamma\psi)(\overline\psi\Gamma\psi)$ has an 
anomalous dimension $\gamma_\Gamma$, and thus transforms under a scale
transformation with parameter $\lambda$ as
\begin{equation}
\label{fftransf}
(\overline\psi\Gamma\psi)(\overline\psi\Gamma\psi)\to
\lambda^{6-\gamma_\Gamma}(\overline\psi\Gamma\psi)(\overline\psi\Gamma\psi)\ ,
\end{equation}
which leads us to introduce a spurion field for $a^2$ transforming as
$a^2\to \lambda^{-2+\gamma_\Gamma}a^2$.   It follows that the operator (\ref{SET}) is
represented in dChPT as
\begin{equation}
\label{EFT}
c_\Gamma a^2{\hat f}_\pi^6 \,e^{(6-\gamma_\Gamma)\tau}{\rm tr}(\Sigma\Gamma\Sigma^\dagger\Gamma)\ ,
\end{equation}
where now $\Gamma$ acts on the taste index of $\Sigma$ \cite{LS,AB,MGLH}, and $c_\Gamma$
is a dimensionless low-energy constant.\footnote{In staggered ChPT both single- and double-trace
operators appear, but only single-trace operators contribute to the taste splittings at leading order.}  Since
there is more than one possible choice for $\Gamma$, we find from Eq.~(\ref{EFT}) that
\begin{equation}
\label{tastesplit}
(aM_\Gamma)^2-(aM_\pi)^2=(a{\hat f}_\pi)^4\sum_{\Gamma'}c_{\Gamma\Gamma'}
e^{(4-\gamma_{\Gamma'})v(m)}\ .
\end{equation}
Assuming, as is the case for QCD, that one operator dominates leads to the simpler
expression
\begin{equation}
\label{tastesplitsimple}
(aM_\Gamma)^2-(aM_\pi)^2=A_\Gamma e^{(4-\gamma_\Gamma)v(m)}\ ,
\end{equation}
and this is the expression we fit to the data, yielding the red and magenta curves in Fig.~2.
The values of the fit parameters are
\begin{eqnarray}
\label{tastefits}
A_{i5}&=&2.0\times 10^{-6}\ ,\qquad \gamma_{i5}=1.9\ ,\\
A_{ij}&=&2.9\times 10^{-6}\ ,\qquad \gamma_{ij}=1.9\ .\nonumber
\end{eqnarray}
As these fits are preliminary, we did not yet estimate errors.   We note, however, that
the fits yield $\gamma_{i5}\simeq\gamma_{ij}\approx 2\gamma_*$.   
A more complete analysis is in preparation.   However,
it is clear that dChPT has a feature not present in standard ChPT: the appearance
of powers of $e^{v(m)}$ in tree-level results.   
These factors explain why the taste splittings are strongly dependent on the fermion
mass to leading order in dChPT at a fixed lattice spacing, in the large-mass regime.   This is 
in sharp contrast with what happens in QCD with
staggered fermions, where taste splittings are independent of the fermion mass to
leading order (and, to a very good precision, in simulations).   The different slopes
seen in Fig.~2 are explained by the different values of $A_{i5}$ and $A_{ij}$ found 
in the fits.

\section{Concluding remarks}
In this preliminary investigation of the LSD data for the 8-flavor $SU(3)$ theory 
using tree-level dChPT, we find that, at least semi-quantitatively, lowest-order
dChPT describes the data quite well.   In particular, dChPT appears to be able
to describe the taste-splitting pattern in staggered discretizations of this theory,
which shows a very different pattern from that of staggered lattice QCD.   We take this 
as a sign that the description of the low-energy behavior of this theory using dChPT
is on the right track.

A complete leading-order analysis of the data is in progress \cite{GNS}.   While this is a 
natural starting point, of course
NLO effects should eventually be considered as well.\footnote{For
some work in this direction, not necessarily within dChPT, see Refs.~\cite{A2,A3,HLS,CM}.}

From our analysis, we can conclude that if indeed dChPT is the correct low-energy
effective theory, the simulations of Refs.~\cite{LSD1,LSD2,LatKMI} are in the ``large-mass''
regime \cite{GS4}, in which the data show approximate hyperscaling.   The intuitive understanding
is that the fermion mass $m$ is large enough to dominate the breaking of scale invariance.   If this is
the case, that would make it more difficult to settle the question whether indeed (as assumed here)
the 8-flavor $SU(3)$ theory is just outside, or already inside, the conformal window.

Given that it is expensive to enlarge the (linear) volume $L$ (while keeping the lattice spacing fixed),
another way of avoiding the large-mass regime would be to make the fermion mass $m$ 
smaller.    This would possibly drive the pions in the theory into the $\epsilon$-regime \cite{LatHC}.   
This regime is also accessible to dChPT \cite{BGKSS}.   However, values of $F_\pi$ would 
be much closer to ${\hat f}_\pi$, which is predicted by dChPT to be very small, and it might 
not be easy to satisfy the requirement that $F_\pi L\gtrsim 1$.

\section*{Acknowledgements}
We would like to thank Ethan Neil for discussions and extensive communication about the data,
and Julius Kuti for discussions.
This work was supported in part by the U.S.\ Department of Energy under
grant DE-SC0013682 (SFSU) and by the Israel Science Foundation under
grant no.~491/17 (Tel Aviv).

\end{document}